\documentstyle{article}
\newcommand{\btok}{$B\to K^* \gamma $~}
\newcommand{\btorho}{$B\to\rho\gamma$~}
\newcommand{\bstok}{$B_s\to K^*\gamma$~}
\newcommand{\bstophi}{$B_s\to \phi \gamma$~}
\newcommand{\btov}{$B\to V \gamma$~}

\begin{document}

\begin{center}{\LARGE \bf Relativistic description of the exclusive
rare \\ radiative decays of $B$~mesons}
\end{center}
\medskip
\begin{center}{R.~N.~Faustov, V.~O.~Galkin\\
\it Russian Academy of Sciences, Scientific Council for
Cybernetics,\\ Vavilov Street 40, Moscow 117333, Russia}
\end{center}
\bigskip
\centerline{\large\bf Abstract}
\smallskip
\noindent  The exclusive rare radiative $B$-decays are studied in  the
framework of the relativistic quark model based on the quasipotential
approach in quantum field theory. The large recoil momentum of the
final vector meson allows for the expansion of the decay form factor
in inverse powers of the b-quark mass.  This considerably simplifies
the analysis of these decays. The $1/m_b$ expansion is carried out up
to the second order. The form factor of \btok decay is found to be
$F_1^{B\to K^* \gamma}(0)=0.32\pm 0.03 $ and it leads to $BR(B\to
K^*\gamma )= (4.5\pm 1.5) \times 10^{-5}$, which is in agreement with
 the recent CLEO data.  The form factors of the decays \btorho,
 \bstophi and \bstok are also considered.  The relation between rare
 radiative and semileptonic $B$-decays into light vector meson is
 discussed.

 \noindent PACS number(s): 12.39Ki, 13.20He, 13.40Hq

 \bigskip
 \section{Introduction}

Rare radiative decays of $B$-mesons represent an important test of
the standard
 model of electroweak interactions. These transitions are induced by
flavour
changing neutral currents and thus they are sensitive probes of new
physics
 (see e.~g. [1]). Such decays are governed by one-loop (penguin)
diagrams with
 the main contribution from virtual top quark and W boson. Therefore,
they
 provide valuable information about the Cabibbo-Kobayashi-Maskawa
(CKM) matrix
 elements $V_{td}$, $V_{ts}$ and $V_{tb}$, and about the top quark
mass. The
first exclusive decay \btok has been observed by CLEO [2]. The
measured
branching ratio is $BR(B\to K^* \gamma)=(4.5 \pm 1.5 \pm 0.9)\times
10^{-5}$.
Also the
inclusive decay rate $B\to X_s\gamma$, obtained from the measurement
of the
photon energy spectrum, has been reported by CLEO: $BR(B\to
X_s\gamma)=
(2.32 \pm 0.51 \pm 0.32 \pm 0.20) \times 10^{-4}$ [3].

Theoretical analysis of rare radiative decays is based on the
effective
Hamiltonian, which is obtained by integrating out the heavy particles.
The renormalization of the effective Hamiltonian coefficients has been
calculated to leading order [4,5]. It turns out that renormalization
effects play an important role. They increase the decay amplitude
approximately
by a factor of two [4]. Some of the next-to-leading order corrections
are
also known [6,7]. The hadronic matrix elements of the effective
Hamiltonian
for inclusive rare radiative decays have been calculated in the
framework
of the heavy quark effective theory (HQET) [8]. It has been shown
that the
leading term in $1/m_b$ expansion corresponds to the parton model
prediction
and the nonperturbative corrections are suppressed by $1/m_b^2$
factor [9,10].
For the calculation of the exclusive decay rates it is necessary to
know the
relevant hadronic form factors. In the case of \btok decay only one
hadronic
form factor $F_1(0)$ contributes. It has been calculated in the
framework of
nonrelativistic quark model [11] and QCD sum rules [12--14]. In [15]
the
heavy quark mass limit has been applied to exclusive \btok decay.
However,
the s-quark in the final $K^*$ meson is not heavy. Its mass is of
order of
$\bar \Lambda$ parameter, which determines the scale of $1/m_Q$
corrections in
HQET [8]. Thus very substantial corrections to this limit come from
the whole
series in $1/m_s$. Nevertheless, the ideas of heavy quark expansion
can be
applied to the exclusive \btok decay. From kinematical analysis it
follows
that the final $K^*$ meson bears large relativistic recoil momentum
$\vert
{\bf\Delta} \vert$ of order of $m_b/2$ and the energy of the same
order.
So it is possible to expand the matrix element of the effective
Hamiltonian
both in inverse powers of b-quark mass from the initial $B$ meson and
in
inverse powers of the recoil momentum $\vert{\bf\Delta} \vert$ of the
final
$K^*$ meson. As a result the expansion in powers of $1/m_b$ arises.
The
aim of this paper is to realize such expansion in the framework of the
relativistic quark model. We show that this expansion considerably
simplifies the analysis of exclusive rare radiative decays of $B$
mesons
and reduce the model dependence of the result. The leading, first and
second order terms of the $1/m_b$ expansion are calculated. It is
necessary to note that rare radiative decays of $B$ mesons require the
completely relativistic treatment, because the recoil momentum of
final
vector meson is highly relativistic.

Our relativistic quark model is based on the quasipotential approach
in
quantum field theory with the specific choice of the $q\bar q$
potential.  It provides a consistent scheme for calculation of all
relativistic corrections at a given order of $v^2/c^2$ and allows for
the heavy quark $1/m_Q$ expansion. This model has been applied for the
calculations of meson mass spectra [16], radiative decay widths [17],
pseudoscalar decay constants [18], semileptonic [19] and nonleptonic
[20] decay rates. The heavy quark $1/m_Q$ expansion in our model for
the heavy-to-heavy semileptonic transitions has been developed in [21]
up to $1/m_Q^2$ order. The results are in agreement with the model
independent predictions of HQET [8]. The rare decay \btok has been
considered in our model in [22]. Here we refine our previous analysis
with more complete account of relativistic effects and using $1/m_b$
expansion. We also consider some other exclusive radiative decays,
including the decay \bstophi and the CKM-suppressed decays \btorho and
$B_s \to K^*\gamma$.

The  paper is organized as follows. In Sect.~2 we briefly review the
theoretical predictions for the inclusive rare radiative $B$-decays
and define the form factor $F_1$, which governs the exclusive decays.
The relativistic quark model is described in Sect.~3, and in Sect.~4
it is applied for the calculation of the rare radiative decay form
factor $F_1$. The $1/m_b$ expansion for this form factor is carried
out in Sect.~5. Our numerical results for the form factors of the
decays \btok, \bstophi, \btorho and \bstok are presented in Sect.~6.
We also discuss the relations between rare radiative and semileptonic
$B$-decays into light vector mesons. Sect.~7 contains our conclusions.

\section{Inclusive and exclusive rare radiative $B$-decays}

In the standard model, $B$-decays are described by the effective
Hamiltonian,
obtained by integrating out the top quark and $W$ boson and using the
Wilson
expansion [4]. For the case of $b \to s$ transition:
\begin{equation}H_{eff}(b\to s)=-\frac{4G_F}{\sqrt2}
V^*_{ts}V_{tb}\sum ^8_{j=1} C_j(\mu)O_j(\mu),\end{equation}
where $V_{ij}$ are the corresponding CKM matrix elements, $ \lbrace
O_j\rbrace $ are a complete set of renormalized dimension six
operators involving light fields, which govern $b\to s$ transitions.
They consist of six four-quark operators $O_j\quad (j=1,\ldots ,6)$,
which determine the non-leptonic $B$-decay rates, the electromagnetic
dipole operator $O_7$
\begin{equation}O_7=\frac{e}{16\pi ^2}\bar s\sigma ^{\mu \nu
}(m_bP_R+m_sP_L)bF_{\mu \nu }, \qquad P_{R,L}=(1 \pm \gamma _5)/2,
\end{equation}
and the chromomagnetic dipole operator $O_8$, which are
responsible for the rare $B$-decays $B \to X_s\gamma$ and $b \to s+g$,
respectively [4]. The Wilson coefficients $C_j(\mu )$ are evaluated
perturbatively at the $W$ scale and then they are evolved down to the
renormalization scale $\mu \sim m_b$ by the renormalization group
equations. The coefficient of the magnetic operator, $C_7$, has been
calculated to leading logarithmic order [4,5].  The next-to-leading
order corrections to the anomalous dimension matrix are also partially
known [6].

The dominant contribution to the inclusive decay width $\Gamma (B\to
X_s\gamma)$
comes from the magnetic moment term $C_7(\mu )O_7(\mu )$. Thus it is
necessary
to calculate the hadronic matrix element of this operator. Recently
it has
been observed that the matrix elements of inclusive decays of hadrons
containing
a heavy quark $Q$ allow for a systematic expansion in powers of
$1/m_Q$ [9].
The leading order term of this expansion reproduces the parton model
rate and the nonperturbative corrections appear only at the second
order of $1/m_Q$ expansion [9]. Therefore, the decay rate for $B\to
X_s \gamma$ is [10]
\begin{eqnarray}\Gamma (B\to X_s\gamma
)&=&\frac{G^2_F\alpha m_b^5}{ 32 \pi ^4} \vert V_{ts}^*V_{tb} \vert
^2 \vert C_7(m_b) \vert ^2\left(1+\frac{m_s^2}{ m_b^2} \right)
\left(1-\frac{m_s^2}{ m_b^2} \right)^3\nonumber\\&  &
\times\left(1+\frac{\lambda _1-9\rho \lambda _2}{
2m_b^2}+O\left(\frac{1}{m_b^3}\right)\right), \end{eqnarray}
with $\rho =(3m_b+5m_s)/( 3m_b-3m_s)$.  The parameter $\lambda _1$
is related to the kinetic energy of the $b$-quark inside $B$ meson and
$\lambda _2$ is related to the $B-B^*$ mass splitting [8].  This rate
is strongly dependent on the value of $b$-quark mass $m_b$, which
depends on the definition. So it is more convinient to connect $\Gamma
(B\to X_s\gamma )$ with the experimentally observed semileptonic decay
rate.

The branching ratio $BR(B\to X_s\gamma )$ can be expressed in terms
of the
inclusive semileptonic branching ratio $BR(B\to X \ell \nu _\ell )$
as [7]
\begin{equation}
BR(B \to X_s \gamma )=6\frac{\alpha}{\pi}\frac{\vert V_{ts}^*
V_{tb}\vert ^2}{\vert V_{cb} \vert ^2}\frac{\vert C_7(m_b)\vert
^2K(m_b)}{ g(m_c/m_b) \big(1-\frac{2}{3}\frac{\alpha_s}{\pi}
f(m_c/m_b)\big)}BR(B\to X\ell \nu _\ell ), \end{equation}
where $g(r)$ is the phase-space factor for $\Gamma (b\to c\ell \nu
_\ell )$:
$g(r)=1-8r^2+8r^6-r^8-24r^4\ln (r)$, and experimental value $BR(B\to
X\ell \nu_\ell )\approx 10.5\%$ is used for the semileptonic branching
fraction.  The function $f(r)$ accounts for QCD corrections to the
semileptonic decay [23] and, for a typical quark mass ratio of
$r=0.35\pm 0.05$, it has the value $f(r)=2.37\mp 0.13$. The factor
$K(m)$ contains $O(\alpha _s)$ corrections to the $B\to X_s\gamma$
rate, due to the QCD bremsstrahlung processes [7].  The resulting
branching ratio is [7,24]
\begin{equation}BR(B\to X_s\gamma )=(3.0\pm 1.2)\times
10^{-4},\end{equation}
for $m_t$ in the range $100<m_t<200$ GeV.

This result agrees with a recent data on the inclusive decay rate by
CLEO
collaboration [3]:
\begin{equation}BR(B\to X_s\gamma )=(2.32\pm 0.51\pm 0.32\pm
0.20)\times 10^{-4}.\end{equation}

In what follows we shall discuss the exclusive rare radiative decays
of $B$
mesons, like $B_{u,d}\to K^*\gamma$, \bstophi and the CKM-suppressed
chanels
$B_{u,d}\to \rho \gamma$, \bstok. We denote all these modes by \btov
, where
$B$ is $B_{u,d}$ or $B_s$ meson and $V$ is $K^*$, $\phi $ or $\rho $.
The
main contribution to the exclusive \btov decay amplitude comes from
the
magnetic moment operator $O_7$. The relevant matrix element has the
covariant
decomposition [11]:
\begin{eqnarray}\langle V(p_V,e)\vert \bar fi\sigma _{\mu \nu}q^\nu
P_Rb\vert B(p_B)\rangle&=&i\epsilon_{\mu \nu \rho \sigma }e^{*\nu
}p_B^\rho p_V^\sigma F_1(q^2)\nonumber\\& & +\big[e_\mu
^*(M_B^2-M_V^2)-(e^*\cdot q)(p_B+p_V)_\mu\big]G_2(q^2),\end{eqnarray}
where $G_2(0)=F_1(0)/2$, $e$ is a polarization vector of final vector
meson,
$q=p_B-p_V$ denotes the four-momenta of the emitted photon. The
exclusive
decay rate is given by
\begin{equation}
\Gamma (B\to K^*\gamma )=\frac{G_F^2\alpha}{ 32\pi ^4}\vert
V_{ts}^*V_{tb} \vert ^2\vert F_1(0)\vert ^2\vert C_7(m_b)\vert
^2(m_b^2+m_s^2)\frac{(M_B^2- M_{K^*}^2)^3}{ M_B^3},\end{equation}
and the analogous expressions can be written for $\Gamma (B_s\to \phi
\gamma )$, $\Gamma (B\to \rho \gamma )$ and $\Gamma (B_s \to K^*\gamma
)$. Then the exclusive branching ratio $BR(B\to K^*\gamma )$ is
\begin{equation}BR(B\to K^*\gamma )=R\cdot BR(B\to X_s\gamma),
\end{equation}
where the ratio of the exclusive to inclusive
radiative decay rates is determined by (3), (8) to be:
\begin{equation}R\equiv\frac{\Gamma (B\to K^*\gamma )}{ \Gamma
(B\to X_s\gamma )}= \frac{\big(1-M_{K^*}^2/M_B^2\big)^3}{
\big(1-m_s^2/m_b^2\big)^3}\frac{M_B^3}{ m_b^3}\frac{\vert F_1(0)\vert
^2}{K(m_b)\left(1+(\lambda_1-9\rho\lambda_2)/(2m_b^2)\right)}.
\end{equation}
The factor $K(m_b)\approx 0.83$ [7], the parameters
$\lambda _1=-0.30\pm 0.30$ GeV$^2$ [8,25] and $\lambda _2=0.12\pm
0.01$ GeV$^2$ [8].

We shall calculate the form factor $F_1(0)$ with the account of
$1/m_b$
corrections up to the second order in the framework of the
relativistic
quark model based on the quasipotential method.

\section{ Relativistic quark model}

In the quasipotential approach meson is described by the wave
function of the bound
quark-antiquark state, which satisfies the quasipotential equation
[26] of
the Schr\"odinger type [27]:
\begin{equation}\left(\frac{b^2(M)}{2\mu_{R}}-\frac{{\bf
p}^2}{2\mu_{R}}\right) \Psi_{M}({\bf p})=\int\frac{d^3 q}{(2\pi)^3}
V({\bf p,q};M)\Psi_{M}({\bf q}),\end{equation}
where the relativistic reduced mass is
\begin{equation}\mu_{R}=\frac{M^4-(m^2_a-
m^2_b)^2}{4M^3};\end{equation}
\begin{equation}b^2(M)=\frac{[M^2-(m_a+m_b)^2]
[M^2-(m_a-m_b)^2]}{4M^2},\end{equation}
$m_{a,b}$ are the quark masses; $ M$ is the meson mass; ${\bf p}$  is
the relative momentum of quarks. While constructing the kernel of this
equation $V({\bf p,q};M)$ --- the quasipotential of quark-antiquark
interaction --- we have assumed that effective interaction is the sum
of the one-gluon exchange term with the mixture of long-range vector
and scalar linear confining potentials. We have also assumed that at
large distances quarks acquire universal nonperturbative anomalous
chromomagnetic moments and thus the vector long-range potential
contains the Pauli interaction. The quasipotential is defined by [16]:
\begin{equation}V({\bf p,q},M)=\bar{u}_a(p)
\bar{u}_b(-p)\Big\{\frac{4}{3}\alpha_SD_{ \mu\nu}({\bf
k})\gamma_a^{\mu}\gamma_b^{\nu}+V^V_{conf}({\bf k})\Gamma_a^{\mu}
\Gamma_{b;\mu}+V^S_{conf}({\bf k})\Big\}u_a(q)u_b(-q),\end{equation}
where $\alpha_S$ is the QCD coupling constant, $D_{\mu\nu}$ is the
gluon propagator; $\gamma_{\mu}$ and $u(p)$ are the Dirac matrices and
spinors; ${\bf k=p-q}$; the effective long-range vector vertex is
\begin{equation}\Gamma_{\mu}({\bf k})=\gamma_{\mu}+
\frac{i\kappa}{2m}\sigma_{\mu\nu}k^{\nu},\end{equation}
$\kappa$ is anomalous chromomagnetic quark moment. Vector and scalar
confining potentials in the nonrelativistic limit reduce to
\begin{equation}V^V_{conf}(r)=(1-\varepsilon)(Ar+B),\quad
V^S_{conf}(r)=\varepsilon(Ar+B),\end{equation}
reproducing $V_{nonrel}^{conf}(r)=V^S_{conf}+V^V_{conf}=Ar+B$,  where
$\varepsilon$ is the mixing coefficient. The explicit expression for
the quasipotential with the account of the relativistic corrections of
order $v^2/c^2$ can be found in ref. [16].  All the parameters of our
model: quark masses, parameters of linear confining potential $A$ and
$B$, mixing coefficient $\varepsilon$ and anomalous chromomagnetic
quark moment $\kappa$ were fixed from the analysis of meson masses
[16] and radiative decays [17].  Quark masses: $m_b=4.88$ GeV;
$m_c=1.55$ GeV; $m_s=0.50$ GeV; $m_{u,d}=0.33$ GeV and parameters of
linear potential: $A=0.18$ GeV$^2$; $B=-0.30$ GeV have standard values
for quark models.  The value of mixing coefficient of vector and
scalar confining potentials $\varepsilon=-0.9$ has been primarily
chosen from the consideration of meson radiative decays, which are
very sensitive to the Lorentz-structure of the confining potential:
the resulting leading relativistic corrections coming from vector and
scalar potentials have opposite signs for the radiative Ml - decays
[17]. Universal anomalous chromomagnetic moment of quark $\kappa=-1$
has been fixed from the analysis of the fine splitting of heavy
quarkonia ${ }^3P_J$- states [16].

Recently we have considered the expansion of the matrix elements of
weak
heavy quark currents between pseudoscalar and vector meson states up
to
the second order in inverse powers of the heavy quark masses [21]. It
has
been found that the general structure of leading, subleading and
second
order $1/m_Q$ corrections in our relativistic model is in accord with
the
predictions of HQET. The heavy quark symmetry and QCD impose rigid
constraints
on the parameters of the long-range potential of our model. The
analysis of the
first order corrections [21] allowed to fix the value of effective
long-range
anomalous chromomagnetic moment of quarks $ \kappa =-1$, which
coincides with
the result, obtained from the mass spectra [16]. The mixing parameter
of vector
and scalar confining potentials has been found from the comparison of
the
second order corrections to be $ \varepsilon =-1$. This value is very
close
to the previous one $\varepsilon =-0.9$ determined from radiative
decays
of mesons [17]. Therefore, we have got QCD and heavy quark symmetry
motivation
for the choice of the main parameters of our model. The found values
of $\varepsilon$ and $\kappa$ imply that confining quark-antiquark
potential has predominantly Lorentz-vector structure, while the scalar
potential is anticonfining and helps to reproduce the initial
nonrelativistic potential.

\section{ Calculation of the rare radiative decay form factor}

The amplitude of the exclusive \btov decay is proportional to the
hadronic
matrix element (7) of the magnetic moment operator $O_7$. Thus it is
necessary
to calculate the transition form factor $F_1(0)$.

The matrix element of the local current $J$ between bound states in
the
quasipotential method has the form [28]:
\begin{equation}\langle V \vert J_\mu (0) \vert B\rangle =\int
\frac{d^3p\, d^3q}{ (2\pi )^6} \bar \Psi_V({\bf p})\Gamma _\mu
({\bf p},{\bf q})\Psi_B({\bf q}),\end{equation}
where $ \Gamma _\mu ({\bf p},{\bf q})$ is the two-particle vertex
function and
$\Psi_{V,B}$ are the meson wave functions projected onto the positive
energy states of quarks.

In the case of rare radiative decays $J_\mu =\bar f i\sigma _{\mu
\nu} k^\nu
P_R b$ and in order to calculate its matrix element between meson
states it
is necessary to consider the contributions to $\Gamma$ from Figs.~1
and 2.
Thus the vertex functions look like
\begin{equation} \Gamma_\mu ^{(1)}({\bf p},{\bf q})=\bar
u_f(p_1)\frac{i}{2}\sigma_{\mu \nu} k^\nu
(1+\gamma^5)u_b(q_1)(2\pi)^3\delta({\bf p}_2-{\bf q}_2),\end{equation}
and
\begin{eqnarray}\Gamma_\mu^{(2)}({\bf p},{\bf q})&=&\bar u_f(p_1)\bar
u_q(p_2)\frac{1}{ 2} \biggl\{ i\sigma_{1\mu
\nu}(1+\gamma_1^5)\frac{\Lambda_b^{(-)}({ k}_1)}{ \varepsilon
_b(k_1)+\varepsilon_b(p_1)}\gamma_1^0V({\bf p}_2-{\bf q}_2)\nonumber\\
& & +V({\bf p}_2-{\bf q}_2)\frac{\Lambda_f^{(-)}(k_1')}{
\varepsilon_f(k_1')+ \varepsilon_f(q_1)}\gamma_1^0i\sigma_{1\mu
\nu}(1+\gamma_1^5)\biggr\}u_b(q_1) u_q(q_2), \end{eqnarray}
where ${\bf k}_1={\bf p}_1-{\bf\Delta};\quad {\bf k}_1'={\bf
q}_1+{\bf\Delta};\quad {\bf\Delta}={\bf p}_B-{\bf p}_V; \quad
\varepsilon (p)=(m^2+{\bf p}^2)^{1/2}$;
$$\Lambda^{(-)}(p)={\varepsilon(p)-\bigl( m\gamma ^0+\gamma^0({\bf
\gamma p})\bigr) \over 2\varepsilon (p)}.$$
Note that the contribution $\Gamma^{(2)}$ is the consequence of the
projection onto the positive-energy states. The form of the
relativistic corrections resulting from the vertex function
$\Gamma^{(2)}$ is explicitly dependent on the Lorentz-structure of
$q\bar q$-interaction.

It is convenient to consider the decay \btov in the $B$ meson rest
frame. Then
the recoil momentum ${\bf\Delta}$ of vector meson $V$ is highly
relativistic
$(\vert {\bf\Delta}\vert \approx M_B/2)$. The wave function of the
meson
moving with the momentum ${\bf\Delta}$ is connected with the wave
function at
rest by the transformation [28]
\begin{equation}\Psi_{V\,{\bf\Delta}}({\bf p})=D_f^{1/2}
(R_{L{\bf\Delta}}^W) D_q^{1/2}(R_{L{ \bf\Delta}}^W)\Psi_{V\,{\bf
0}}({\bf p}),\end{equation}
where $D^{1/2}(R)$ is the well-known rotation matrix and $R^W$  is the
Wigner rotation.

The meson wave functions in the rest frame have been calculated by
numerical
solution of the quasipotential equation (11) [29]. However, it is more
convenient to use analytical expressions for meson wave functions. The
examination of numerical results for the ground state wave functions
of mesons
containing at least one light quark has shown that they can be well
approximated
by the Gaussian functions
\begin{equation}\Psi_M({\bf p})\equiv \Psi_{M\,{\bf 0}}({\bf
p})=\Bigl(\frac{4\pi}{ \beta_M^2} \Bigr)^{3/4}\exp\Bigl(-\frac{{\bf
p}^2}{ 2\beta_M^2}\Bigr),\end{equation}
with the deviation less than 5\%.

The parameters are

$$\beta_B=0.41\ {\rm GeV};\qquad\beta_{K^*}=0.33\ {\rm GeV};
\qquad\beta_\rho=0.31\ {\rm GeV};$$
\begin{equation}\beta_{B_s}=0.46\ {\rm GeV};\qquad\beta_\phi=0.36\
{\rm GeV}.\end{equation}

Substituting the vertex functions (18), (19), with the account of
wave function
transformation of vector meson (20), in the matrix element (17) we
get for
the form factor $F_1$ the following expression:
\begin{eqnarray}F_1(0)&=&F_1^{(1)}(0)+\varepsilon
F_1^{S(2)}(0)+(1-\varepsilon)F_1^{V(2)}(0), \\
F_1^{(1)}(0)&=&\sqrt{\frac{E_V}{ M_B}}\int \frac{d^3p}{(2\pi)^3}
\bar \Psi_V\left({\bf p}+\frac{2\varepsilon_q}{
E_V+M_V}{\bf\Delta}\right) \sqrt{\frac{\varepsilon_f(p+\Delta) +m_f}
{ 2\varepsilon_f(p+\Delta)}}\sqrt{\frac{\varepsilon_b (p)+m_b}{
2\varepsilon_b(p)}}\nonumber\\
& &\times\Biggl\{ 1+\frac{p_z^2+({\bf p\Delta})}{(\varepsilon_f
(p+\Delta)+ m_f) (\varepsilon_b(p)+m_b)}+(M_B-E_V)\Biggl[\frac{1}
{\varepsilon_f(p+\Delta)+m_f}\nonumber\\
& &+\frac{({\bf p\Delta})}{{\bf
\Delta}^2}\left(\frac{1}{\varepsilon_f(p+\Delta)+m_f}+
\frac{1}{\varepsilon_b(p)+m_b}\right)\nonumber\\
& &-\frac{p_x^2+p_y^2}{ 2(E_V+M_V)}\Biggl(\frac{1}{
\varepsilon_q(p)+m_q}\left(\frac{1}{ \varepsilon_b(p)+m_b}
-\frac{1}{ \varepsilon_f(p+\Delta)+m_f}\right)\nonumber\\
& &+\frac{4}{ (\varepsilon_f(p) +m_f)(\varepsilon_b(p)+m_b)}
\Biggr)\Biggr]\Biggr\}\Psi_B({\bf p}),\\
F_1^{S(2)}(0)&=&\sqrt{\frac{E_V}{ M_B}} \int\frac{d^3p}{(2\pi)^3}
\bar\Psi_V\left({\bf p}+\frac{2\varepsilon_q}{
E_V+M_V}{\bf\Delta}\right) \sqrt\frac{\varepsilon_f(p+\Delta)+m_f}{
2\varepsilon_f(p+\Delta)}\nonumber\\
& &\times \Biggl\{\frac{\varepsilon_f(\Delta)-
m_f}{2\varepsilon_f(\Delta)
(\varepsilon_f(\Delta)+m_f)}\left(1+\frac{M_B-E_V}{ 2m_b}\frac{({\bf
p\Delta})}{{\bf \Delta}^2}\right)\nonumber\\
& &\times\left(M_V-\varepsilon_f\left(p+\frac{2\varepsilon_q}{
E_V+M_V}\Delta\right)- \varepsilon_q\left(p+\frac{2\varepsilon_q}{
E_V+M_V}\Delta\right)\right)-(M_B-E_V)\frac{({\bf p\Delta})}{
{\bf\Delta}^2}\nonumber\\
& &\times\Biggl(\frac{M_B+M_V-\varepsilon_b(p)-\varepsilon_q(p)-
\varepsilon_f\left(p+\frac{2\varepsilon_q}{E_V+M_V}\Delta\right)
-\varepsilon_q\left(p+\frac{2\varepsilon_q}{E_V+M_V}\Delta\right)}{
2m_b(\varepsilon_b(\Delta)+m_b)}\nonumber\\
& &+\frac{M_B-M_V-\varepsilon_b(p)-\varepsilon_q(p)+
\varepsilon_f\left(p+\frac{2\varepsilon _q}{E_V+M_V}
\Delta\right)+\varepsilon_q\left(p+\frac{2\varepsilon_q}{
E_V+M_V}\Delta\right) }{2\varepsilon_f(\Delta)(\varepsilon_f(\Delta)
+m_f)}\Biggr)\Biggr\} \Psi_B({\bf p}),\\
F_1^{V(2)}(0)&=&\sqrt{\frac{E_V}{ M_V}}\int\frac{d^3p}{ (2\pi)^3}
\bar\Psi_V\left({\bf p}+\frac{2\varepsilon_q}{
E_V+M_V}{\bf\Delta}\right) \sqrt{\frac{\varepsilon_f(p+\Delta)+m_f}{
2\varepsilon_f(p+\Delta)}}\nonumber\\
& &\times \Biggl\{\frac{\varepsilon_f(\Delta)-
m_f}{2\varepsilon_f(\Delta)
(\varepsilon_f(\Delta)+m_f)}\left(1+\frac{M_B-E_V}{ 2m_b}\frac{({\bf
p\Delta})}{{\bf\Delta}^2}\right)\nonumber\\
& &\times\left(M_V-\varepsilon_f\left(p+\frac{2\varepsilon_q}{
E_V+M_V}\Delta\right)- \varepsilon_q\left(p+\frac{2\varepsilon_q}{
E_V+M_V}\Delta\right)\right)+(M_B-E_V)\frac{({\bf p\Delta})}{
{\bf\Delta}^2}\nonumber\\
& &\times\Biggl(\frac{M_V-\varepsilon_f\left(p+{2\varepsilon_q}{
E_V+M_V}\Delta\right) -\varepsilon_q\left(p+\frac{2\varepsilon_q}{
E_V+M_V}\Delta\right)}{ 2m_b(\varepsilon_b(\Delta)+m_b)}
+\frac{M_B-\varepsilon_b(p)-\varepsilon_q(p)}{
2\varepsilon_f(\Delta) (\varepsilon_f(\Delta)+m_f)}\nonumber\\
& &+\frac{1+\kappa}{2(\varepsilon_q(p)+m_q)}\left(
\frac{1}{\varepsilon_b(\Delta) +m_b}-\frac{1}{
\varepsilon_f(\Delta)}\right)\biggr(M_B-M_V-
\varepsilon_b(p)\nonumber\\
& &+\varepsilon_f\left(p+\frac{2\varepsilon_q}{
E_V+M_V}\Delta\right)+\varepsilon_q \left(p+\frac{2\varepsilon_q}{
E_V+M_V}\Delta\right)\biggr)\Biggr)\Biggr\}\Psi_B({\bf p}),
\end{eqnarray}
where the superscripts ``(1)" and ``(2)" correspond to Figs.~1 and 2,
$S$ and $V$ --- to the scalar and vector potentials of $q\bar
q$-interaction;
\begin{equation}\vert {\bf\Delta}\vert=\frac{M_B^2-M_V^2}{
2M_B};\qquad E_V=\frac{M_B^2+M_V^2}{ 2M_B}; \end{equation}
and $z$-axis is chosen in the direction of ${\bf\Delta}$. The
contributions to the form factor coming from Fig.~2 are proportional
to the binding energy of $B$ or $V$ mesons. Taking into account that
in our model $m_b+m_d\simeq M_b$; $m_b+m_s\simeq M_{B_s}$;
$m_s+m_d\simeq M_{K^*}$; $2m_s\simeq M_\phi$; $2m_d\simeq M_\rho$ and
$\vert{\bf\Delta}\vert^2\simeq m_b^2/4\gg \langle {\bf p}^2\rangle$ we
have neglected the terms proportional to the product of binding
energies and ratios ${\bf p}^2/\varepsilon_f^3(\Delta)$, ${\bf p}^2/
\varepsilon_b^3(\Delta)$. The omitted terms are of order $1/m_b^3$.

The expressions (23)--(26) for the form factor $F_1(0)$ differ from
our
previous results [22] in the argument of the wave function of final
vector
meson $V$. In [22] we have used for simplicity the nonrelativistic
limit
${\bf p}+{m_q\over M_V}{\bf \Delta}$ for this argument, taking into
account
the approximation of the wave functions by Gaussians. However, our
recent
analysis of the $1/m_Q$ corrections to the semileptonic decays [21]
has
shown, that such approximation is not completely adequate. It is
necessary
to take relativistic expression ${\bf p}+{2\varepsilon_q\over E_V+M_V}
{\bf\Delta}$ in the argument of the wave function, what is done in
eqs.(23)--(26).

\section{ $1/m_b$ expansion for the form factor $F_1(0)$ }

At present considerable theoretical interest is attracted to HQET.
This theory is based on QCD and heavy quark expansion. Additional
heavy quark symmetries (for a review see e.~g. [8]) arise in the limit
of infinitely heavy quark mass. HQET provides a systematic expansion
in inverse powers of the heavy quark mass of hadronic matrix elements
between mesons with one heavy and one light quarks. In the case of
exclusive heavy-to-heavy transitions heavy quark symmetries
considerably reduce the number of independent form factors in each
order of $1/m_Q$ expansion. This simplifies the theoretical analysis
of such decays. In exclusive heavy-to-light meson transitions (such as
the \btov decays) the predictive power of HQET is essentialy less.
Only the relations between various form factors of the semileptonic
and rare $B$-decays are imposed in the limit of infinitely heavy
$b$-quark mass [30]. However, as it has been already mentioned in the
introduction, the large value of the recoil momentum of final vector
meson $\vert {\bf\Delta}\vert \sim m_b/2$ allows the expansion in
powers of $1/m_b$ for the radiative rare decay form factor $F_1(0)$.
This expansion leads to considerable simplification of the formulae
for $F_1(0)$. It has been already partly used by us in derivation of
(23)--(26). The large value of recoil momentum $\vert {\bf
\Delta}\vert$
allowed us to neglect ${\bf p}^2$ in comparison with ${\bf
\Delta}^2$ in the quark energy $\varepsilon_f(p+\Delta)$ in final
meson in the expressions for $F_1^{S(2)}$ and $F_1^{V(2)}$. Thus we
were able to perform one of the integrations in the current matrix
element (17) using quasipotential equation as in the case of the heavy
final meson [17,19,28]. As a result, we get more compact formulae. In
this section for the sake of consistency we carry out the complete
expansion of (23)--(26) in inverse powers of $b$-quark mass.

In HQET the mass of $B$ meson has the following expansion in $1/m_b$
[8]
\begin{equation}M_B=m_b+\bar\Lambda+\frac{\Delta m_B^2}{
2m_b}+O\left(\frac{1}{ m_b^2}\right),\end{equation}
where parameter $\bar\Lambda$ is the difference between the meson and
quark masses in the limit of infinitely heavy quark mass. In our model
$\bar\Lambda $ is equal to the mean value of light quark energy inside
the heavy meson $\bar\Lambda=\langle\varepsilon_q\rangle_B\approx
0.54$~GeV [21]. $\Delta m_B^2$ arises from the first-order power
corrections to the HQET Lagrangian and has the form [8]:
\begin{equation}\Delta m_B^2=-\lambda_1-3\lambda_2.\end{equation}
The parameter $\lambda_1$ results from the mass shift due to the
kinetic operator, while $\lambda_2$ parameterizes the chromomagnetic
interaction [8]. The value of spin-symmetry breaking parameter
$\lambda_2$ is related to the vector-pseudoscalar mass splitting
$$\lambda_2\approx {1\over 4}(M_{B^*}^2-M_B^2)=0.12\pm 0.01\ {\rm
GeV}^2.$$
The parameter $\lambda_1$ is not directly connected with
observable quantities. Theoretical predictions for it vary in a wide
range:  $\lambda_1=-0.30\pm 0.30 \ {\rm GeV}^2$ [8,25].

In the limit $m_Q\to \infty$, meson wave functions become independent
of the flavour of heavy quark. Thus the Gaussian parameter $\beta_B$
in (21) should have the following expansion [21]
\begin{equation}\beta_B=\beta-\frac{\Delta\beta^2}{
m_b}+O\left(\frac{1}{ m_b^2}\right),\qquad \beta \approx  0.42 \
{\rm GeV},\end{equation}
where the second term breaks the flavour symmetry and in our model is
equal to $\Delta\beta^2\approx 0.045\ {\rm GeV}^2$ [21].

Substituting (28) in (27) we get the $1/m_b$ expansion of the recoil
momentum and the energy of final vector meson:
\begin{eqnarray}\vert{\bf\Delta}\vert&=&\frac{m_b}{
2}\left(1+\frac{1}{m_b}\bar\Lambda +\frac{1}{ m_b^2}\left(\frac{\Delta
m_B^2}{ 2}-M_V^2\right)\right)+O\left(\frac{1}{m_b^2}\right)
,\nonumber\\
E_V&=&\frac{m_b}{ 2}\left(1+\frac{1}{ m_b}\bar \Lambda+\frac{1}{
m_b^2}\left(\frac{\Delta m_B^2}{2}
+M_V^2\right)\right)+O\left(\frac{1}{
m_b^2}\right).\end{eqnarray}

Now we use the Gaussian approximation for the wave functions (21).
Then shifting the integration variable ${\bf p}$ in (23)--(26) by
$-\frac{\varepsilon_q}{ E_V+M_V}{\bf\Delta}$, we can factor out
the ${\bf\Delta}$ dependence of the meson wave function overlap in
form factor $F_1$. The result can be written in the form
\begin{equation}F_1(0)={\cal F}_1({\bf\Delta}^2)\exp(-\zeta {\bf
\Delta}^2),\end{equation}
where $\vert{\bf\Delta}\vert$ is given by (27) and
\begin{equation}\zeta{\bf\Delta}^2=\frac{2\tilde\Lambda^2{\bf\Delta}^2
(\beta_B^2+\beta_V^2)(E_V+M_V)^2}=\frac{2\tilde\Lambda^2}{
(\beta_B^2 +\beta_V^2)}\left(\frac{M_B-M_V}{
M_B+M_V}\right)^2,\end{equation}
here $\tilde\Lambda$ is equal to the mean value of light quark energy
between heavy and light meson states:
\begin{equation}\tilde\Lambda
=\langle\varepsilon_q\rangle.\end{equation}
Expanding (33) in powers of $1/m_b$ we get
\begin{equation}\zeta{\bf\Delta}^2=\frac{\tilde\Lambda^2}{
\beta^2}\eta\left(1-4{M_V\over m_b}\right)+O\left(\frac{1}{
m_b^2}\right),\end{equation}
where $\eta=\frac{2\beta_B^2}{ \beta_B^2+\beta_V^2}$. We see that
the first term in this expansion is large. Really, even in the case of
the lightest final meson $\rho$: $4{M_\rho / m_b}\approx 0.63$.
The value of this correction to the form factor $F_1(0)$ is also
increased by the exponentiating in (32). Therefore, we conclude that
the first order correction in $1/m_b$ expansion of $F_1(0)$, arising
from the meson wave function overlap, is large. Thus, in the
following, we use unexpanded expression (33) in the exponential of the
form factor $F_1(0)$ in (32).

In contrast to the meson wave function overlap the factor ${\cal
F}_1({\bf \Delta}^2) $ in (32) has a well defined $1/m_b$ expansion.
First and second order corrections are small. Substituting the
Gaussian wave functions (21) in the expressions for the form factor
(23)--(26), with the value of anomalous chromomagnetic quark moment
$\kappa=-1$, and using (32) and the expansions (28)--(31), we get up
to the second order in $1/m_b$ expansion:
\begin{eqnarray}{\cal F}_1({\bf\Delta}^2)&=&{\cal
F}_1^{(1)}({\bf\Delta}^2)+ \varepsilon{\cal
F}_1^{S(2)}({\bf\Delta}^2)+(1-\varepsilon){\cal F}_1^{V(2)}({\bf
\Delta}^2);\\
{\cal F}_1^{(1)}({\bf\Delta})&=&N\biggl(1-\frac{1}{
2m_b}\left(\tilde \Lambda\eta-\left\langle {\bf p}^2\left(\frac{1}{
2}\frac{1}{\bar\varepsilon_q +m_q}-\frac{1}{3}\frac{1}{
\bar\varepsilon_f+m_f}\right)\right\rangle\right)\nonumber\\
& & +\frac{1}{ 2m_b^2}\biggl(M_V^2-m_f^2-\frac{13}{
4}\langle{\bf p}^2 \rangle-\frac{1}{4}\tilde\Lambda^2\eta^2
+\tilde\Lambda\eta(4m_f+2M_V)\nonumber\\
& &+\tilde\Lambda\eta^2\frac{\beta_V^2}{\beta^2}\frac{\Delta\beta^2}{
\beta}-\frac{1}{3}\left\langle\frac{{\bf p}^2}{
\bar\varepsilon_q+m_q}\right\rangle \left(\frac{5}{
2}m_f+3M_V+2\bar\Lambda -4\tilde\Lambda\eta\right)\nonumber\\
& & +\frac{1}{ 3}\left\langle\frac{{\bf p}^2}{\bar\varepsilon_f
+m_f}\right\rangle (2M_V-m_f)\biggr)\biggr);\\
{\cal F}_1^{S(2)}({\bf \Delta}^2)&=&N\frac{1}{
2}\biggl(\frac{1}{ m_b}\left(1-\frac{1}{m_b}(3m_f+\bar\Lambda+
\frac{1}{ 2}\tilde\Lambda\eta)\right)(M_V-
\langle\bar\varepsilon_q\rangle
-\langle\bar\varepsilon_f\rangle)\nonumber\\
& & +\frac{\tilde\Lambda\eta}{ m_b^2}\biggl(\frac{1}{
2+\sqrt5}(\bar \Lambda+M_V-\langle\bar\varepsilon_f\rangle-
2\langle\bar\varepsilon_q \rangle)+2(\bar\Lambda-M_V+\langle\bar
\varepsilon_f\rangle)\nonumber\\
& & -\frac{3}{ 2}\left\langle\frac{{\bf p}^2}{ \bar
\varepsilon_q}\right\rangle - \frac{1}{ 3}\left(\frac{5}{
 2}-\frac{1}{ 2+\sqrt5} \right)\left\langle\frac{{\bf p}^2}{
 \bar\varepsilon_f}\right\rangle\biggr) \biggr);\\
{\cal F}_1^{V(2)}({\bf\Delta}^2)&=&N\frac{1}{
2}\biggl(\frac{1}{ m_b}\left(1-\frac{1}{ m_b}(3m_f+ \bar\Lambda
+\frac{1}{ 2}\tilde\Lambda\eta)\right)(M_V-\langle
\bar\varepsilon_q\rangle -\langle\bar\varepsilon_f\rangle)\nonumber\\
& & -\frac{\tilde\Lambda\eta}{ m_b^2}\biggl(\frac{1}{
2+\sqrt5}(M_V- \langle\bar\varepsilon_f\rangle-\langle
\bar\varepsilon_q\rangle)+2(\bar\Lambda-\langle
\bar\varepsilon_q\rangle)\nonumber\\
& & +\frac{1}{6}\left(1+\frac{2}{ 2+\sqrt5}\right)
\left\langle\frac{{\bf p}^2}{\bar\varepsilon_f}\right\rangle-
\left(\frac{1}{2}- \frac{1}{ 3(2+\sqrt5)} \right)\left\langle
\frac{{\bf p}^2}{\bar\varepsilon_q}\right\rangle\biggr)\biggr),
\end{eqnarray}
where $N=\left(\frac{2\beta_B\beta_V}{
\beta_B^2+\beta_V^2}\right)^{3/2}
 =\left(\frac{\beta_V}{\beta_B}\eta\right)^{3/2}$ is due to the
 normalization of Gaussian wave functions in (21); $\bar\varepsilon_i
 =\sqrt{{\bf p}^2+m_i^2+\tilde\Lambda^2\eta^2}\quad (i=f,q)$, i.~e.
 the energies of light quarks in final vector meson acquire additional
 contribution from the recoil momentum. The averaging is taken over
 the Gaussian wave functions of $B$ and $V$ mesons, so it can be
 carried out analytically. For example,
\begin{equation}\langle\bar\varepsilon_i\rangle=\frac{1}{
 \sqrt\pi}\frac{\bar m_i^2}{\beta_V\sqrt\eta}e^zK_1(z), \end{equation}
 where $\bar m_i^2=m_i^2+\tilde\Lambda^2\eta^2$ and $K_1(z)$ is the
 modified Bessel function; $z={\bar m_i^2/ 2\eta\beta_V^2}$.
 Analogous expressions can be obtained for the other matrix elements
 in (37)--(39).

 Our final result for the rare radiative decay \btov form factor
 $F_1(0)$ is given up to the second order of $1/m_b$ expansion by
 eqs.~(32), (33) and (36)--(39).

 \section{ Results and discussion }

 Using the parameters (22) of the wave functions (21) in the
 expressions for the form factor $F_1(0)$ (32),(33) and (36)--(39) we
 get (for the values of the anomalous chromomagnetic quark moment
 $\kappa=-1$ and the mixing parameter of vector and scalar confining
 potentials $\varepsilon=-1 $ [21]):
 \begin{eqnarray}F_1^{B\to  K^*\gamma}(0)&=&0.32\pm
 0.03\qquad F_1^{B\to \rho \gamma}(0)\ =\ 0.26\pm 0.03\nonumber\\
 F_1^{B_s\to\phi\gamma}(0)&=&0.27\pm 0.03\qquad F_1^{B_s\to
 K^*\gamma}(0)\ =\ 0.23\pm 0.02.\end{eqnarray}
 The theoretical uncertainty in (41) results mostly from the
 approximation of the wave functions by Gaussians (21) and does not
 exceed 10\% of form factor values. The contributions of higher order
 terms in $1/m_b$ expansion in (36)--(39) are negligibly small. Thus
 the unexpanded (23)--(26) and expanded (41) values of form factors
 differ unessentially. This conclusion is confirmed by numerical
 analysis.

 Our results (41) for the form factor values are in good agreement
 with recent calculations within the light-cone QCD sum rule [13] and
 hybrid sum rule [14] approaches. The comparison of predictions is
 given in Table 1. All results agree within errors.

 The calculated value of form factor $F_1^{B\to K^*\gamma}$ in (41)
 yields for the ratio (10) of exclusive to inclusive radiative decay
 widths:
\begin{equation}R(B\to K^*\gamma)\equiv\frac{\Gamma (B\to
 K^*\gamma)}{\Gamma (B\to X_s\gamma)}=(15\pm 3)\%.\end{equation}
 Combining this result with the QCD-improved inclusive radiative
 branching ratio $BR(B\to X_s\gamma)$ given by (5), we get
\begin{equation}BR^{th}(B\to K^*\gamma)=(4.5\pm 1.5)\times
 10^{-5}.\end{equation}
 This branching ratio agrees well with experimental measurement by
 CLEO [2]
 $$BR^{exp}(B\to K^*\gamma)=(4.5\pm 1.5\pm 0.9)\times
 10^{-5}.$$

 Our numerical analysis has shown that $1/m_b$ corrections in
 (37)--(39) give rather small contributions to the decay form factor.
 So in the first crude approximation we can neglect them. This
 corresponds to the limit $m_b\to\infty$ for ${\cal F}_1({\bf
 \Delta}^2)$. Then the dependence on the Lorentz structure of
 confining potential is lost and we get a simple formula
\begin{equation} F_1(0)=\frac{M_B+M_V}{ 2\sqrt{M_B
 M_V}}\tilde\xi(w),\end{equation}
 with $w=\frac{M_B^2+M_V^2}{ 2M_B M_V}$. We have introduced the
 function
\begin{equation}\tilde\xi(w)=\left(\frac{\beta_V}{
 \beta}\eta\right)^{3/2} \left(\frac{2}{ w+1}\right)^{1/2}
 \exp\left(-\eta\frac{\tilde\Lambda^2}{\beta^2}\frac{w-1}{
 w+1}\right),\end{equation}
 which in the limit of infinitely heavy final quark in $V$ meson,
 coinsides with the Isgur-Wise function of our model [21]:
\begin{equation}\xi(w)=\left(\frac{2}{
 w+1}\right)^{1/2}\exp\left(-\frac{\bar\Lambda^2}{
 \beta^2}\frac{w-1}{ w+1}\right),\qquad w\equiv v\cdot v'=
\frac{M_B^2
 +M_V^2-q^2}{ 2M_B M_V}.\end{equation}
 Thus if we take the formal limit of infinitely heavy $s$-quark, as
it
 is done in [15], we get for \btok form factor
\begin{equation}F_1(0)=\frac{M_B+M_{K^*}}{ 2\sqrt{M_B
 M_{K^*}}}\xi(w),\quad {\rm for}\ \ m_b\to\infty,\
 m_s\to\infty,\end{equation}
 with $w=\frac{M_B^2+M_{K^*}^2}{ 2M_B M_{K^*}}$. The relation
 (47) is, certainly, not adequate, because $s$-quark is not heavy. It
 overestimates the form factor approximately by a factor of 1.5.

 The approximate formula (44) gives the values of form factors, which
 are slightly higher than (41). However, the difference does not
 exceed 10\%. Thus we can conclude that the $q^2$-dependence of form
 factor $F_1$ near $q^2=0$ is determined by the function (45).

 We can use our results for $F_1$ to test the HQET relations [30]
 between the form factors of rare radiative and semileptonic decays of
 $B$ mesons. Isgur and Wise [30] have shown that in the limit of
 infinitely heavy $b$-quark mass an exact relation connects the form
 factor $F_1$ with the semileptonic decay form factors defined by:
\begin{eqnarray}\langle V(p_V,e)\vert \bar f\gamma_\mu(1-\gamma_5)b
 \vert B(p_B)\rangle&=&-(M_B+M_V)A_1(q^2)e_\mu^*\nonumber \\
 & & +\frac{2V(q^2)}{M_B+M_V}i\epsilon_{\mu\nu\rho\sigma} e^{*\nu}
 p_B^\rho p_V^\sigma+\ldots\end{eqnarray}
 where the ellipses denote terms proportional to $(p_B+p_V)_\mu$ or
$q_\mu$. This relation is valid for $q^2$ values sufficiently close
to
$q_{max}^2=(M_B-M_V)^2$ and reads:
\begin{equation}F_1(q^2)=\frac{q^2+M_B^2-M_V^2}{
2M_B}\frac{V(q^2)}{ M_B+M_V}+ \frac{M_B+M_V}{2M_B}A_1(q^2).
\end{equation}
It has been argued in [31,32,14], that in these processes the soft
contributions dominate over the hard perturbative ones, and thus the
Isgur-Wise relation (49) could be extended to the whole range of
$q^2$.

We can calculate the form factors of semileptonic $B$-decay in vector
light meson at the $q^2=0$ point using the method developed here for
rare radiative decay form factor $F_1(0)$. It is easy to see that the
$1/m_b$ corrections coming from the overlap of meson wave functions
are also large. The similar exponential factor as in eq.~(32) arises,
and we shall not expand it in the $1/m_b$. The $1/m_b$ corrections to
the preexponential factor are again rather small. So here we limit our
analysis only to the leading order of $1/m_b$ expansion for this
factor. In this approximation the semileptonic decay form factors are
independent of the Lorentz-properties of the confining $q\bar
q$-potential and are given [33] by a simple formulae
\begin{eqnarray}V(0)&=&\frac{M_B+M_V}{2\sqrt{M_BM_V}}\tilde\xi(w),\\
A_1(0)&=&\frac{2\sqrt{M_B M_V}}{ M_B+M_V}\frac{1}{
2}(1+w)\tilde\xi(w),\end{eqnarray}
where $\tilde\xi(w)$ is defined by (45). In the limit of infinitely
heavy $b$- and $f$-quarks the form factors (50) and (51) satisfy all
the relations imposed by HQET [8]. The $q^2$-dependence of the form
factors near $q^2=0$ is determined for $V_1$ by the function $\tilde
\xi(w)$ and for $A_1$ by the product $\frac{1}{ 2}(1+w)\tilde\xi(w)$.

It is now easy to check that form factors (50), (51) and (32) satisfy
Isgur-Wise relation (49) for the $q^2$ values near zero. The complete
analysis of $1/m_b$ corrections to the semileptonic $B$-decays into
light mesons up to the second order terms will be given elsewhere
[33].

\section{ Conclusions}

We have investigated the rare radiative decays of $B$ mesons. The
large value ($\sim m_b/2$) of the recoil momentum of the final vector
meson requires the completely relativistic treatment of these decays.
On the other hand, the presence of large recoil momentum in the energy
of the final meson allows for the $1/m_b$ expansion of weak decay
matrix element. The contributions to this expansion come both from
heavy $b$-quark mass and large recoil momentum of light vector meson.

Using the quasipotential approach in quantum field theory and the
relativistic quark model, we have performed the $1/m_b$ expansion of
the rare radiative $B$-decay form factor $F_1(0)$ up to the second
order. We have found that $1/m_b$ corrections coming from the overlap
of meson wave functions are large. This agrees with the results of the
hybrid sum rules [14], where considerable $1/m_b$ corrections for the
form factor $F_1(0)$ have been also obtained. Thus we treat the ${\bf
\Delta}^2$-dependence of $F_1$, arising from the overlap of meson wave
functions, without expansion. All other $1/m_b$ corrections turn out
to be small. They reduce the value of the form factor $F_1(0)$
approximately by 10\%. We have calculated the form factors for the
decays \btok, \btorho and \bstophi, \bstok. The found values (41) are
in good agreement with the predictions of QCD sum rules [13,14].
Combining our result for $F_1^{B\to K^*\gamma}(0)=0.32\pm 0.03$ with
the QCD-improved theoretical prediction for inclusive branching ratio
(5), we obtain $BR(B\to K^*\gamma)=(4.5\pm 1.5)\times 10^{-5}$, which
is in accord with the experimental branching ratio $BR(B\to K^*\gamma)
=(4.5\pm 1.5\pm 0.9)\times 10^{-5}$ [2].

We have also evaluated [33] the form factors of $B$ meson semileptonic
decays into light vector mesons at the point of maximal recoil. It has
been found that the relations between rare and semileptonic decay form
factors [30], obtained in the limit of infinitely heavy $b$-quark, are
satisfied in our model.

\smallskip
\bigskip
\noindent {\large \bf Aknowledgements }
\nopagebreak
\smallskip

We express our gratitude to B.~A.~Arbuzov, M.~A.~Ivanov,
J.~G.~K\"orner, V.~A.~Matveev, M.~Neubert, V.~I.~Savrin for the
interest in our work and helpful discussions of the results. We are
also grateful to A.~Yu.~Mishurov for the help in numerical
calculations and G.~G.~Likhachev for the help in preparation of this
paper for publication. This research was supported in part by the
Russian Foundation for Fundamental Research under Grant
No.94-02-03300-a and Interregional Centre for Advanced Studies.

\bigskip
\bigskip

\noindent {\large \bf References}

\frenchspacing
\begin{enumerate}
\item J. L. Hewett, SLAC preprint SLAC-PUB-6521 (1994); R.
Barbieri and G. F. Giudice, Phys. Lett. B 309 (1993) 86; L. Randall
and R.  Sundrum, Phys. Lett. B 312 (1993) 148; H. Anlauf, SLAC
preprint SLAC-PUB-6525 (1994); F. M. Borzumati, Z. Phys. C 63 (1994)
291
\item R.Ammar et al. (CLEO Coll.), Phys. Rev. Lett. 71 (1993) 674
\item B. Barish et al. (CLEO Coll.), Contribution to the
International
Conference on High Energy Physics, Glasgow, July 20--27, 1994;
preprint CLEO CONF 94-1 (1994)
\item B. Grinstein, R. Springer and M. Wise, Phys. Lett. B 202 (1988)
138; Nucl. Phys. B 339 (1990) 269; R. Grigjanis, P. J. O'Donnel, M.
Sutherland and H. Navelet, Phys. Lett. B 237 (1990) 252; G. Cella, G.

Curci, G. Ricciardi and A. Vicere, Phys. Lett. B 248 (1990) 181; B
325
(1994) 227; M. Misiak: Phys. Lett. B 269 (1991) 161; B 321 (1994)
193;
Nucl. Phys. B 393 (1993) 23
\item M. Ciuchini et al., Phys. Lett. B 316 (1993) 127; Nucl.  Phys.
B
421 (1994) 41
\item A. J. Buras et al., Nucl. Phys. B 370 (1992) 69;
B 400 (1993) 37; B 400 (1993) 75; M. Ciuchini et al., Phys. Lett. B
301 (1993) 263; Nucl. Phys. B 415 (1994) 403
\item A. Ali and C. Greub, Z. Phys. C 60 (1993) 433; Phys. Lett. B
287
(1992) 191
\item M. Neubert, Phys. Rep. 245 (1994) 259
\item J. Chay, H. Georgi and B. Grinstein, Phys. Lett. B 247
(1990) 399; I. I. Bigi, N. G. Uraltsev and A. I. Vainshtein, Phys.
Lett. B 293 (1992) 430; A. F. Falk, M. Luke and M. J. Savage, Phys.
Rev. D 49 (1994) 3367
\item M. Neubert, CERN preprint CERN-TH.7113/93 (1993)
\item T. Altomari, Phys. Rev. D 37 (1988) 677; P. J. O'Donnell and
H. K. K. Tung, Phys. Rev. D 44 (1991) 741; N. G. Deshpande et al., Z.

Phys. C 40 (1988) 369
\item T. M. Aliev, A. A. Ovchinnikov and V. A. Slobodenyuk, Phys.
Lett. B 237 (1990) 569; C. A. Dominguez, N. Paver and Riazuddin,
Phys.
Lett. B 214 (1988) 459; P. Colangelo et al., Phys. Lett. B 317 (1993)
183
\item A. Ali, V. M. Braun and H. Simma, Z. Phys. C 63 (1994) 437
\item S. Narison, CERN preprint CERN-TH.7166/94 (1994)
\item A. Ali and T. Mannel, Phys. Lett. B 264 (1991) 447; A. Ali, T.
Ohl and T. Mannel, Phys. Lett. B 298 (1993) 195
\item V. O. Galkin, A. Yu. Mishurov and R. N. Faustov, Yad. Fiz.
55 (1992) 2175
\item V. O. Galkin and R. N. Faustov, Yad. Fiz. 44 (1986) 1575;
V.  O. Galkin, A. Yu. Mishurov and R. N. Faustov, Yad. Fiz. 51 (1990)
1101
\item V. O. Galkin, A. Yu. Mishurov and R. N. Faustov, Yad.
Fiz. 53 (1991) 1676
\item V. O. Galkin, A. Yu. Mishurov and R. N. Faustov, Yad. Fiz. 55
(1992) 1080
\item R. N. Faustov, V. O. Galkin and A. Yu. Mishurov,
Proc. Seventh Int. Seminar ``Quarks'92", eds. D. Yu. Grigoriev et al.
(World Scientific, Singapore 1993), p.~326
\item R. N. Faustov and V. O. Galkin, Z. Phys. C in press
\item R. N. Faustov and V. O. Galkin, Mod. Phys. Lett. A 7 (1992)
2111
\item A. Ali and E. Pietarinen, Nucl. Phys. B 154 (1979) 519; G.
Altarelli et al., Nucl. Phys. B 208 (1982) 365
\item A. Ali, CERN preprint CERN-TH.7168/94 (1994)
\item T. Mannel, CERN preprint CERN-TH.7162/94 (1994); Z. Ligeti and
Y. Nir, Phys. Rev. D 49 (1994) R4331; I. Bigi et al., CERN preprint
CERN-TH.7361/94 (1994); M. Neubert, CERN preprint CERN-TH.7395/94
(1994); P. Ball and V. M. Braun, Phys. Rev. D 49 (1994) 2472
\item A. A. Logunov and A. N. Tavkhelidze, Nuovo Cimento 29 (1963)
380
\item A. P. Martynenko and R. N. Faustov, Teor. Mat. Fiz. 64
(1985) 179
\item R. N. Faustov, Ann. Phys. 78 (1973) 176; Nuovo
Cimento 69 (1970) 37
\item V. O. Galkin and R. N. Faustov, Teor. Mat. Fiz. 85 (1990) 155
\item N. Isgur and M. B. Wise, Phys. Rev. D 42 (1990) 2388
\item G. Burdman and J. F. Donoghue, Phys. Lett. B 270 (1991) 55
\item S. Narison, CERN preprint CERN-TH.7237/94 (1994)
\item R. N. Faustov, V. O. Galkin and A. Yu. Mishurov, in preparation
(hep-ph/9505321)
\end{enumerate}
\nonfrenchspacing
\vfill
\newpage

\noindent {\Large\bf Table caption}

\smallskip
\noindent {\bf Table 1}
Comparison of our predictions for the rare radiative decay form
factor
$F_1(0)$ with light-cone QCD sum rule [13] and hybrid sum rule [14]
results.

\bigskip
\noindent {\Large \bf Figure captions}

\smallskip
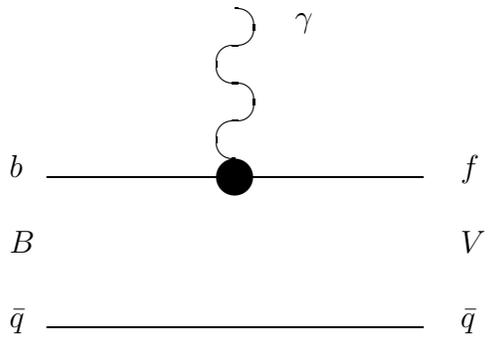
\noindent {\bf Fig.~1} Lowest order vertex function

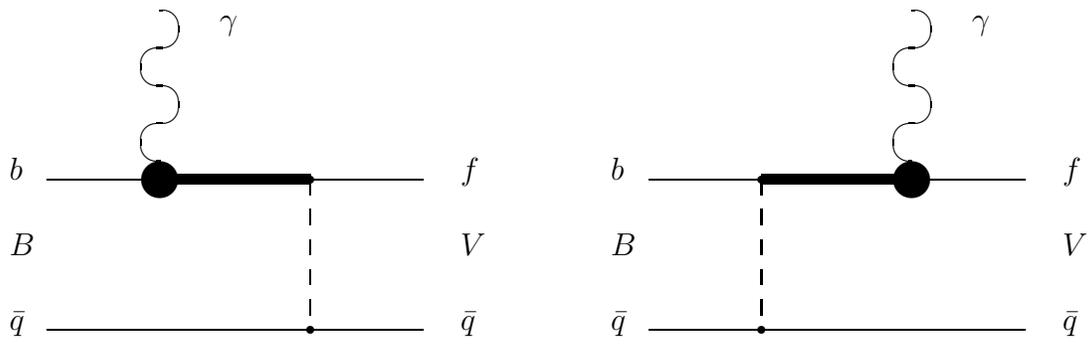
\noindent {\bf Fig.~2} Vertex function with the account of the quark
interaction. Dashed line corresponds to the effective potential (14).
Bold line denotes the negative-energy part of quark propagator.

\vfill

{\large \bf Table 1}

\bigskip
\begin{tabular}{cccc}
\hline
\hline
Decay & our results& {[13]} & {[14]} \\
\hline
$B\to K^*\gamma$ & $0.32\pm0.03$ &
$0.32\pm0.05$ & $0.308\pm0.013\pm0.036\pm0.006$ \\
$B\to\rho\gamma$ & $0.26\pm0.03$ &
$0.24\pm0.04$ & $0.27\pm0.011\pm0.032$\\
$B_s\to\phi\gamma$ & $0.27\pm0.03$ &
$0.29\pm0.05$ & \\
$B_s\to K^*\gamma$ & $0.23\pm0.02$ & $0.20\pm0.04$ & \\
\hline
\hline
\end{tabular}

\hfill
\vfill
\newpage
\unitlength=1mm
\begin{picture}(150,150)
\put(10,100){\line(1,0){50}}
\put(10,120){\line(1,0){50}}
\put(35,120){\circle*{5}}
\multiput(32.5,130)(0,-10){2}{\begin{picture}(5,10)
\put(2.5,10){\oval(5,5)[r]}
\put(2.5,5){\oval(5,5)[l]}\end{picture}}
\put(5,120){\large$b$}
\put(5,100){\large$\bar q$}
\put(5,110){\large$B$}
\put(65,120){\large$f$}
\put(65,100){\large$\bar q$}
\put(65,110){\large$V$}
\put(43,140){\large$\gamma$}
\put(30,85){\Large\bf Fig. 1}
\put(10,20){\line(1,0){50}}
\put(10,40){\line(1,0){50}}
\put(25,40){\circle*{5}}
\put(25,40){\thicklines \line(1,0){20}}
\multiput(25,40.5)(0,-0.1){10}{\thicklines \line(1,0){20}}
\put(25,39.5){\thicklines \line(1,0){20}}
\put(45,40){\circle*{1}}
\put(45,20){\circle*{1}}
\multiput(45,40)(0,-4){5}{\line(0,-1){2}}
\multiput(22.5,50)(0,-10){2}{\begin{picture}(5,10)
\put(2.5,10){\oval(5,5)[r]}
\put(2.5,5){\oval(5,5)[l]}\end{picture}}
\put(5,40){\large$b$}
\put(5,20){\large$\bar q$}
\put(5,30){\large$B$}
\put(65,40){\large$f$}
\put(65,20){\large$\bar q$}
\put(65,30){\large$V$}
\put(33,60){\large$\gamma$}
\put(90,20){\line(1,0){50}}
\put(90,40){\line(1,0){50}}
\put(125,40){\circle*{5}}
\put(105,40){\thicklines \line(1,0){20}}
\multiput(105,40.5)(0,-0.1){10}{\thicklines \line(1,0){20}}
\put(105,39,5){\thicklines \line(1,0){20}}
\put(105,40){\circle*{1}}
\put(105,20){\circle*{1}}
\multiput(105,40)(0,-4){5}{\line(0,-1){2}}
\multiput(122.5,50)(0,-10){2}{\begin{picture}(5,10)
\put(2.5,10){\oval(5,5)[r]}
\put(2.5,5){\oval(5,5)[l]}\end{picture}}
\put(85,40){\large$b$}
\put(85,20){\large$\bar q$}
\put(85,30){\large$B$}
\put(145,40){\large$f$}
\put(145,20){\large$\bar q$}
\put(145,30){\large$V$}
\put(133,60){\large$\gamma$}
\put(70,5){\Large \bf Fig. 2}

\end{picture}

\end{document}